\documentclass[aip,apl,reprint,superscriptaddress,citeautoscript]{revtex4-1}
\usepackage{graphicx}%
\usepackage{dcolumn}%
\usepackage{bm}%

\begin{document}

\author{H. Peelaers}
  \email{peelaers@ku.edu}
  \affiliation{Materials Department, University of California, Santa Barbara, California 93106-5050, USA}
  \affiliation{Department of Physics and Astronomy, University of Kansas, Lawrence, Kansas 66045, USA}
\author{E. Kioupakis}
\affiliation{Department of Materials Science and Engineering, University of Michigan, Ann Arbor, Michigan 48109, USA}
\author{C. G. Van de Walle}
\affiliation{Materials Department, University of California, Santa Barbara, California 93106-5050, USA}

\date{\today}

\title{Limitations of In$_2$O$_3$ as a transparent conducting oxide}

\begin{abstract}
Sn-doped In$_2$O$_3$ or ITO is the most widely used transparent conducting oxide.
We use first-principles calculations to investigate the limitations to its transparency
due to free-carrier absorption mediated by phonons or charged defects.
We find that the main contribution to the phonon-assisted indirect absorption is due to emission (as opposed to absorption) of phonons,
which explains why the process is relatively insensitive to temperature. The wavelength dependence of this indirect absorption process can be described by a power law.
Indirect absorption mediated by charged defects or impurities is also unavoidable since doping is required to obtain conductivity.
At high carrier concentrations, screening by the free carriers becomes important.
We find that charged-impurity-assisted absorption becomes larger than phonon-assisted absorption for impurity concentrations above 10$^{20}$ cm$^{-3}$. The differences in the photon-energy dependence of the two processes can be explained by band-structure effects.
\end{abstract}

\maketitle

In$_2$O$_3$, and in particular Sn-doped In$_2$O$_3$ (usually referred to as ITO), is the most widely used transparent conducting oxide (TCO)~\cite{Ellmer2012}. The material combines transparency to visible light with high conductivity, allowing for a wide range of applications, such as transparent electrodes in flat-panel displays~\cite{Katayama1999} or solar cells~\cite{Fortunato2007}, (opto)electronic devices~\cite{Bierwagen2015}, IR-reflective window coatings~\cite{Hamberg1986}, plasmonics~\cite{Liu2014}, and integration with Ga$_2$O$_3$ electronics~\cite{Peelaers2015b}.

The fundamental band gap of In$_2$O$_3$ is around 2.6-2.9 eV~\cite{Weiher1966,Janowitz2011,Scherer2012,King2009,Walsh2008,Morris2018}; however, strong optical absorption starts only around 3.5-3.7 eV~\cite{Weiher1966,Janowitz2011,Scherer2012,King2009,Walsh2008,Morris2018}.
The absence of absorption from valence to conduction band is a necessary but not sufficient condition for transparency to visible light.
Achieving high conductivity requires introducing a high concentration of electrons in the conduction band; carrier concentrations as high as 2x10$^{21}$ cm$^{-3}$ have been reported~\cite{Bierwagen2014}.
Excitation of these free carriers to higher-energy states can also lead to optical absorption; a fundamental study of this process is the subject of this Letter.

\begin{figure}[tb]
	\centering
	\includegraphics[width=\columnwidth]{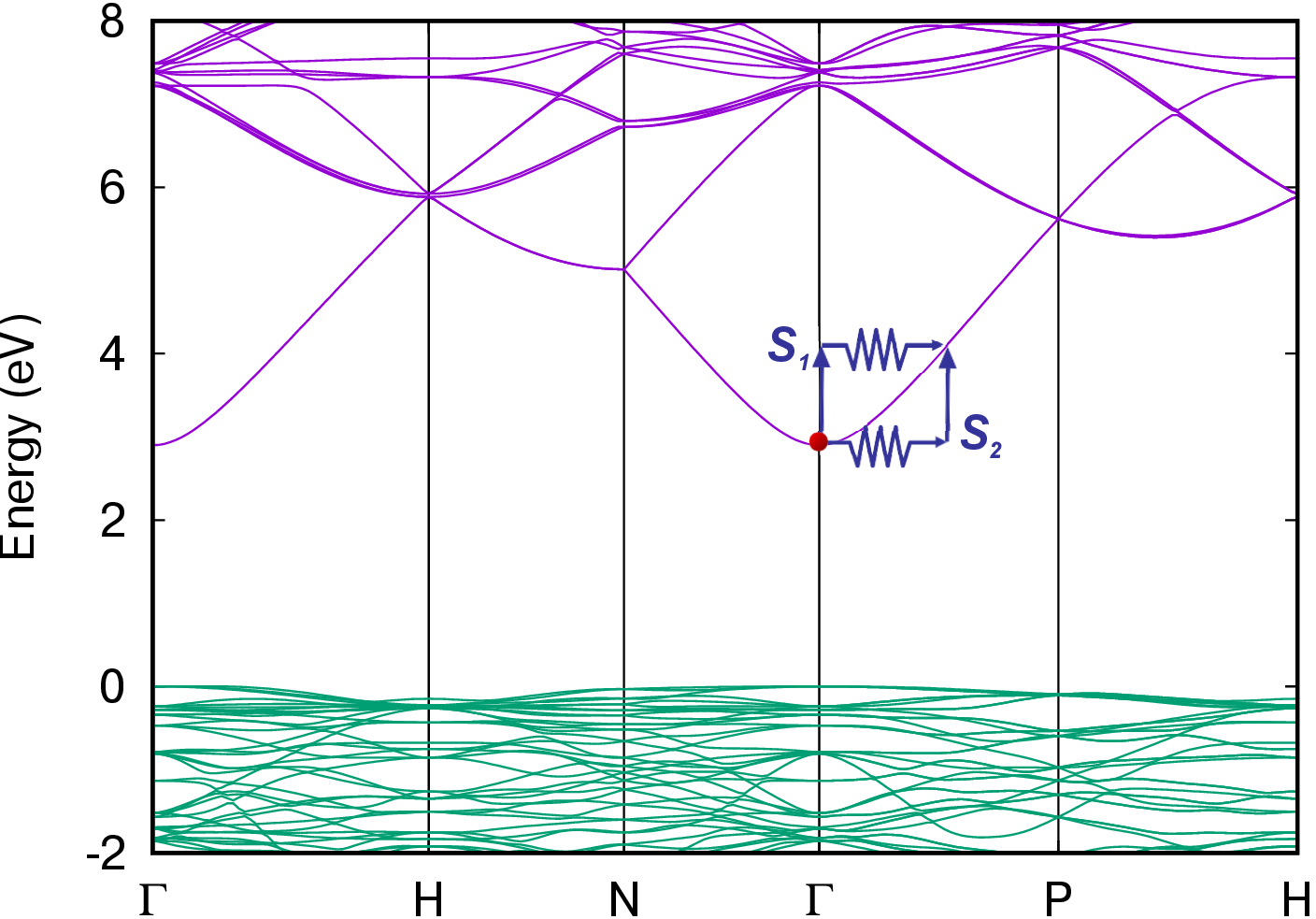}
	\caption{(Color online) The electronic band structure of In$_2$O$_3$ along high-symmetry {\bf k}-point paths. The zero energy is set at the valence-band maximum. The transitions corresponding to the $\bm{S}_1$ and $\bm{S}_2$ matrix elements [Eq.~(\ref{eq:S})] are illustrated.}
	\label{fig:bandplot}
\end{figure}

Direct transitions of free carriers in the conduction band to higher-lying conduction bands (see Fig.~\ref{fig:bandplot}) are not possible with visible-light photons;
only indirect transitions can lead to absorption within the visible range.
Such processes are usually described by a Drude model; here we will use first-principles calculations, i.e., without any fitting parameters, to go beyond such a phenomenological approach. This allows us to describe the fundamental limitations to the transparency of In$_2$O$_3$ caused by the interactions of phonons and charged impurities (or defects) with the free carriers.
We will first discuss the phonon-assisted process, followed by a discussion of the effect of charged impurities, which are unavoidably present due to the need for large concentrations of free carriers to obtain good conductivity. For the phonon-assisted process we compare our results, based on first-principles calculations of matrix elements, with the commonly used Fr\"ohlich approximation, and we discuss why absorption in In$_2$O$_3$ is much weaker than in SnO$_2$.  We explicitly include screening, for both the phonon-assisted and charged-impurity processes.

Our calculations are performed using density functional theory within the local-density approximation (LDA)~\cite{Ceperley1980,*Perdew1981}. We used norm-conserving Troullier-Martins pseudopotentials~\cite{Troullier1991} in the \textsc{Quantum-ESPRESSO}~\cite{Giannozzi2009} package, with a cutoff energy of 90 Ry for the plane-wave basis. The Brillouin zone is sampled with a 4x4x4 Monkhorst-Pack~\cite{Monkhorst1976} {\bf k}-point mesh. Phonons and electron-phonon coupling matrix elements are calculated within density functional perturbation theory~\cite{Baroni2001} on a 24$\times$24$\times$24 {\bf q}-grid. %

\begin{figure}[t]
	\centering
	\includegraphics[width=0.95\columnwidth]{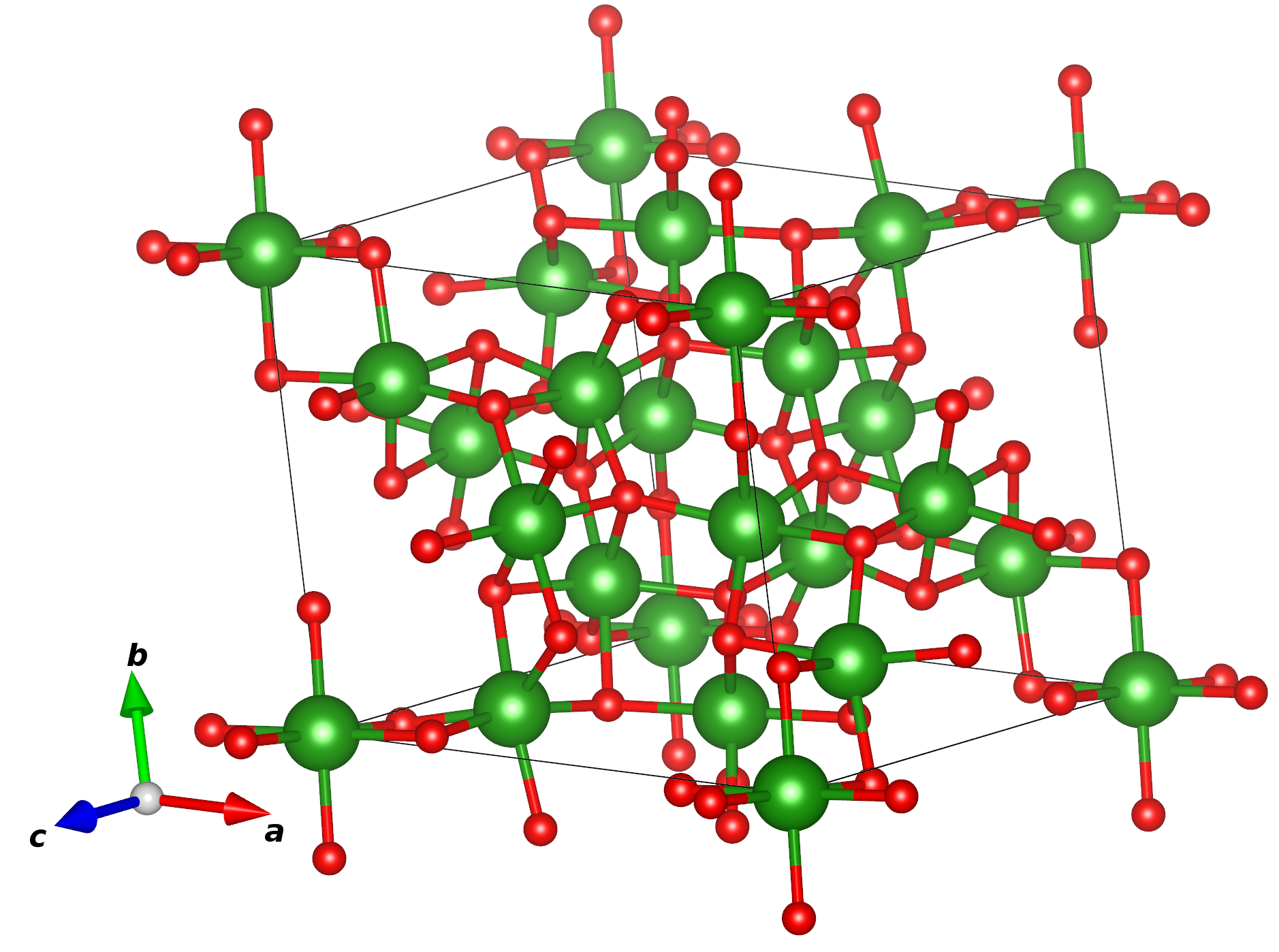}
	\caption{(Color online) The 40-atom unit cell of bixbyite In$_2$O$_3$. The large spheres are In atoms and the smaller spheres O atoms. Indium atoms occupy distorted octahedral positions on either the Wyckoff $24d$ or $8d$ sites, and oxygen atoms occupy the $48e$ positions. }
	\label{fig:structure}
\end{figure}

In$_2$O$_3$ can occur in several polymorphs~\cite{Fuchs2008}; here we focus on the most stable structure, which is cubic bixbyite, with space group \#206 or $Ia\bar{3}$. Its unit cell, shown in Fig.~\ref{fig:structure}~\cite{Momma2011}, consists of 40 atoms.
We obtain a lattice parameter of 10.15 \AA, in good agreement with the experimental value of 10.12 \AA~\cite{Marezio1966}.
Since there are 40 atoms in the unit cell, 120 phonon modes are present, which significantly increases the computational burden of calculating the electron-phonon interactions on a fine {\bf q}-point mesh.
Our calculated frequencies at the $\Gamma$ point agree well with previous calculations~\cite{Garcia-Domene2012} and with experiment~\cite{Garcia-Domene2012,Kranert2014,White1972,Korotcenkov2005,MateiGhimbeu2008,Zhang2007,Berengue2010,Rojas-Lopez2000}. A comparison of the Raman-active phonon frequencies is shown in the Supplementary Material in Table S1.

The electronic band structure is shown in Fig.~\ref{fig:bandplot}.  Since LDA underestimates the band gap, a scissor shift is applied to the conduction bands to reproduce the experimental band gap. However, since we are only considering free-carrier absorption here, the exact value of the band gap is not important; what matters is the conduction-band structure. To ascertain the validity of the LDA conduction-band structure, we compared it with a calculation performed with a hybrid functional~\cite{Heyd2006}. The energy and dispersion of conduction bands were very similar in the two calculations, with differences that would have negligible effects on the calculated free-carrier absorption [see Fig. S1 of the Supplementary Material]. 
Figure~\ref{fig:bandplot} shows that the conduction-band minimum is located at the $\Gamma$ point.
The valence bands have low dispersion and mainly O $p$ character, while the lowest conduction band is comprised of $s$ orbitals and has a large dispersion.
The electron effective mass calculated from LDA, 0.16 $m_e$, is close to the experimental value of 0.18 $m_e$~\cite{Feneberg2016}.

Phonons mediate absorption by providing the momentum necessary to reach unoccupied conduction-band states away from the zone center. We describe this indirect absorption process using Fermi's golden rule~\cite{Peelaers2015}:
\begin{eqnarray}
\label{eq:alpha2}
\alpha(\omega) = & 2\frac{4 \pi^2 e^2}{\omega c n_r(\omega)} \frac{1}{V_{\text{cell}}} \frac{1}{N_{\bm{k}} N_{\bm{q}}} \sum_{\nu i j \bm{k} \bm{q}}\left| \bm{\hat{e}} \cdot \left( \bm{S}_1+\bm{S}_2\right) \right| ^2 \nonumber \\
 & \times P  \delta(\epsilon_{j,\bm{k}+\bm{q}}-\epsilon_{i\bm{k}}-\hbar\omega \pm \hbar\omega_{\nu \bm{q}} ) \, .
\end{eqnarray}
$V_{\text{cell}}$ is the unit-cell volume, $\hbar\omega$ and $\bm{\hat{e}}$ are the energy and polarization of the absorbed photon, $n_r(\omega)$ the refractive index of In$_2$O$_3$ at frequency $\omega$, $\hbar\omega_{\nu \bm{q}}$ the phonon energy, and $\epsilon_{i\bm{k}}$ the electron energy. $i$ and $j$ indicate the band number, $\nu$ the phonon mode, $\bm{k}$ and $\bm{q}$ the wave vectors, and $N_{\bm{k}}$ and $N_{\bm{q}}$ are the number of $\bm{k}$ and $\bm{q}$ wave vectors in the grid. The generalized optical matrix elements $\bm{S}_1$ and $\bm{S}_2$ are given by
\begin{eqnarray}
 \bm{S}_1(\bm{k},\bm{q}) = & \sum_m \frac{
\bm{v}_{im}(\bm{k}) g^{\text{el-ph}}_{mj,\nu}(\bm{k},\bm{q})
}{\epsilon_{m\bm{k}}-\epsilon_{i\bm{k}}-\hbar\omega}, \nonumber\\
 \bm{S}_2(\bm{k},\bm{q}) = & \sum_m\frac{
g^{\text{el-ph}}_{im,\nu}(\bm{k},\bm{q}) \bm{v}_{mj}(\bm{k}+\bm{q})
}{\epsilon_{m,
\bm{k}+\bm{q}}-\epsilon_{i\bm{k}}\pm \hbar\omega_{\nu \bm{q}} } \label{eq:S},
\end{eqnarray}
and correspond to the two possible paths of the indirect absorption process (see Fig.~\ref{fig:bandplot}). $\bm{v}_{im}(\bm{k})$ are the optical matrix elements and $g^{\text{el-ph}}_{mj,\nu}(\bm{k},\bm{q})$ the electron-phonon coupling matrix elements.
The factor $P$ accounts for the carrier and phonon statistics and contains the temperature dependence,
\begin{equation}
\label{eq:P}
P = \left(n_{\nu \bm{q}} +\frac{1}{2} \pm \frac{1}{2}
\right)(f_{i\bm{k}}-f_{j,\bm{k}+\bm{q}}).
\end{equation}
Here $n_{\nu \bm{q}}$ and $f_{i\bm{k}}$ are the phonon and electron occupation numbers. The upper (lower) sign corresponds to phonon emission (absorption).
Note that all quantities entering here are calculated from first principles, without any fitting parameters. Eq.(\ref{eq:alpha2}) contains a sum over both {\bf k}- and {\bf q}-points. By making the assumption that all free carriers are located near the $\Gamma$ point, and that these have similar electron-phonon matrix elements, we can replace the double sum by a single sum over {\bf q}-points, which is essential for rendering the computations tractable.
The assumption that the carriers are located near the $\Gamma$ point is fully justified for electron concentrations up to $3\times10^{18}$ cm$^{-3}$, while the assumption for the matrix elements is valid for Fermi levels up to at least 0.5 eV above the conduction-band minimum (corresponding to carrier concentrations up to $2\times10^{20}$ cm$^{-3}$). To obtain results that are independent of the free-carrier concentration, we will report the absorption cross section, which is the absorption coefficient $\alpha$ divided by the free-carrier concentration.

\begin{figure}[tb]
	\centering
	\includegraphics[width=\columnwidth]{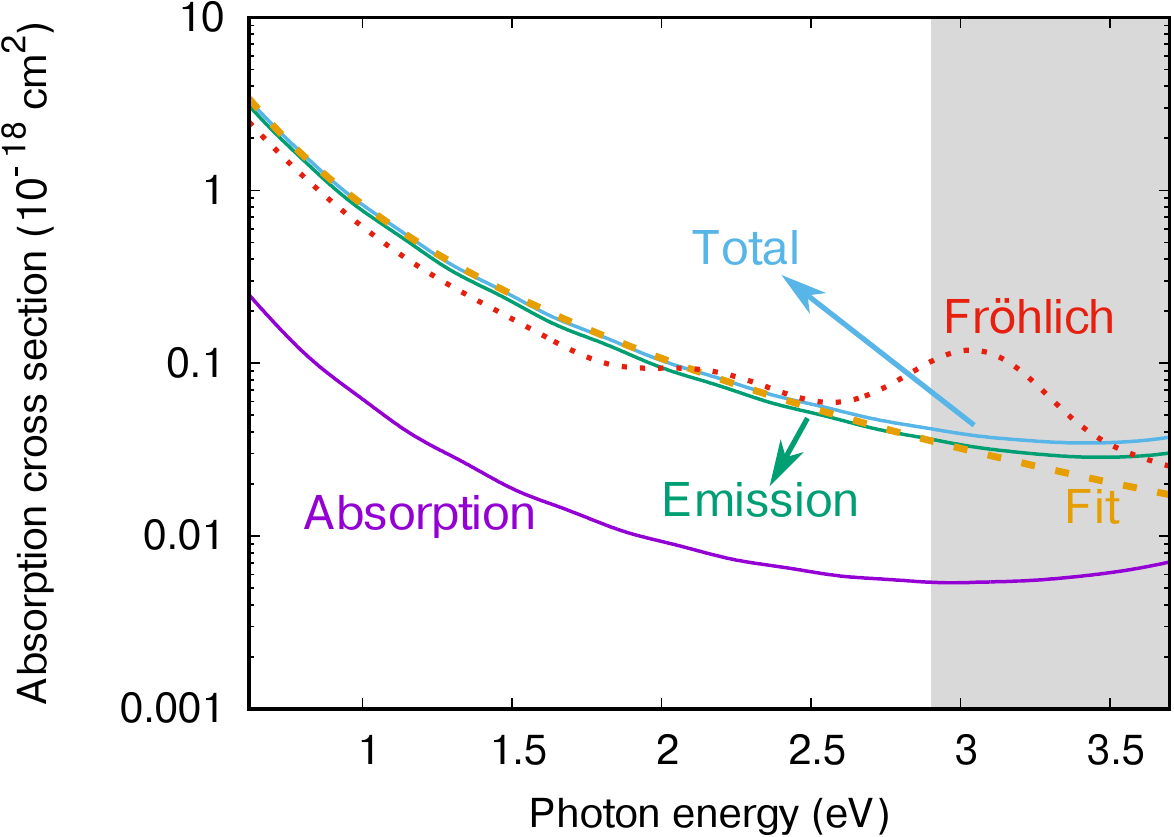}
	\caption{(Color online) Calculated phonon-assisted absorption cross section for In$_2$O$_3$ (at 300 K) as a function of photon energy. The grey area indicates energies for which weak across-the-gap transitions can take place. Contributions from phonon absorption {\it versus} phonon emission are identified. The dashed line indicates a fit to a power law, as described in the text.  The dotted line indicates the result obtained by the Fr\"ohlich model.}
	\label{fig:indabs}
\end{figure}

There are two possible processes: either an existing phonon can be absorbed or a new phonon can be emitted. Our results (at 300 K) are shown in Fig.~\ref{fig:indabs} for photon energies up to 3.7 eV, the energy at which strong absorption from valence to conduction bands sets in.
However, weak across-the-gap absorption has been observed with an onset at the fundamental gap; \cite{Weiher1966,Irmscher2014} therefore we shade the energy range between 2.9 and 3.7 eV in grey.
The phonon emission process is clearly the dominant absorption process at 300 K.  This also implies that lowering the temperature would not significantly affect phonon-assisted absorption, since phonons can be emitted even at 0 K. Fig. S2 shows a comparison between 0 K and 300 K.

The indirect absorption increases with decreasing photon energy (or longer photon wavelength). We find that the relation between phonon-assisted indirect absorption and photon energy can be expressed by a power law. In Fig.~\ref{fig:indabs} we show a fit to $\sigma/(10^{-18}\text{cm}^2)=a\times (\hbar\omega/\text{eV})^{-3}$, where we fitted in the region of photon energies from 0.62 eV to 2.7 eV.
The fit yields a coefficient $a$ = 0.83. The agreement between the first-principles results and the fit is very good at low photon energies; deviations are observed only for photon energies larger than 2.5 eV. This fit implies that the absorption is inversely proportional to the cube power of the photon energy, or equivalently proportional to the cube of the photon wavelength. Such a relationship is expected in the case of linear dispersion (a good approximation for conduction bands that show strong nonparabolicity) and absorption dominated by longitudinal-optical (LO) phonons~\cite{Peelaers2012}.

Our first-principles results in Fig.~\ref{fig:indabs} take all phonon modes and all possible transitions between electronic bands into account.
It is informative to compare these results with a simplified model, in which we assume that the main contribution to phonon-assisted absorption is due to LO phonon modes, and that only the highest phonon mode contributes. The electron-phonon coupling matrix elements are then given by the Fr\"ohlich model
\begin{equation}\label{eq:frohlich}
g_F(\bm{q})=\frac{1}{q}\sqrt{\frac{2\pi\hbar\omega_{\text{LO}}e^2}{V_{\text{cell}}}\left(\frac{1}{\epsilon_\infty}-\frac{1}{\epsilon_0}\right)} \,
,
\end{equation}
where $\omega_{\text{LO}}$ is the frequency of the LO mode at the $\Gamma$ point and $\epsilon_\infty$ and $\epsilon_0$ are the high-frequency and static dielectric constants.
For photon energies up to about 2 eV, where intraband transitions dominate, the single-LO phonon mode Fr\"ohlich model closely approximates the full first-principles results.  The slight underestimation is related to the fact that not all modes are taken into account. Nevertheless, the good agreement indicates that LO phonon modes, with a $1/q$ dependence of the electron-phonon matrix elements, are dominant.

The absorption cross section calculated by the Fr\"ohlich model shows a slight upturn at energies just below 2 eV, which is caused by interband processes: an inspection of the band structure (Fig.~\ref{fig:bandplot}) shows that at energies of about 2 eV above the conduction-band minimum the first conduction band becomes degenerate with the second conduction band, for example along the segment H-N. However, the Fr\"ohlich model overestimates the magnitude of these interband electron-phonon matrix elements, which is even more evident for interband transitions around 3 eV.

The Fr\"ohlich model also provides insight into why phonon-assisted indirect absorption in In$_2$O$_3$ is about 50\% smaller compared to SnO$_2$,~\cite{Peelaers2012,Peelaers2015} another TCO material.
Based on Eq.~(\ref{eq:frohlich}), we see that electron-phonon coupling is stronger in SnO$_2$.
SnO$_2$ is more ionic than In$_2$O$_3$, i.e., there is a larger difference between the static and high-frequency dielectric constants and hence the factor $\left(\frac{1}{\epsilon_\infty}-\frac{1}{\epsilon_0}\right)$ in the Fr\"ohlich model [Eq.~(\ref{eq:frohlich})] is larger in SnO$_2$.
In addition, the highest LO phonon frequency in SnO$_2$ is larger than the highest LO frequency in In$_2$O$_3$.
Both effects contribute to SnO$_2$ having stronger absorption than In$_2$O$_3$.
However, an even larger role is played by the density of states [which enters Eq.~(\ref{eq:alpha2}) through the energy-conserving delta function] being larger in SnO$_2$ than in In$_2$O$_3$, due to the smaller effective mass for electrons of In$_2$O$_3$.

\begin{figure}[tb]
	\centering
	\includegraphics[width=\columnwidth]{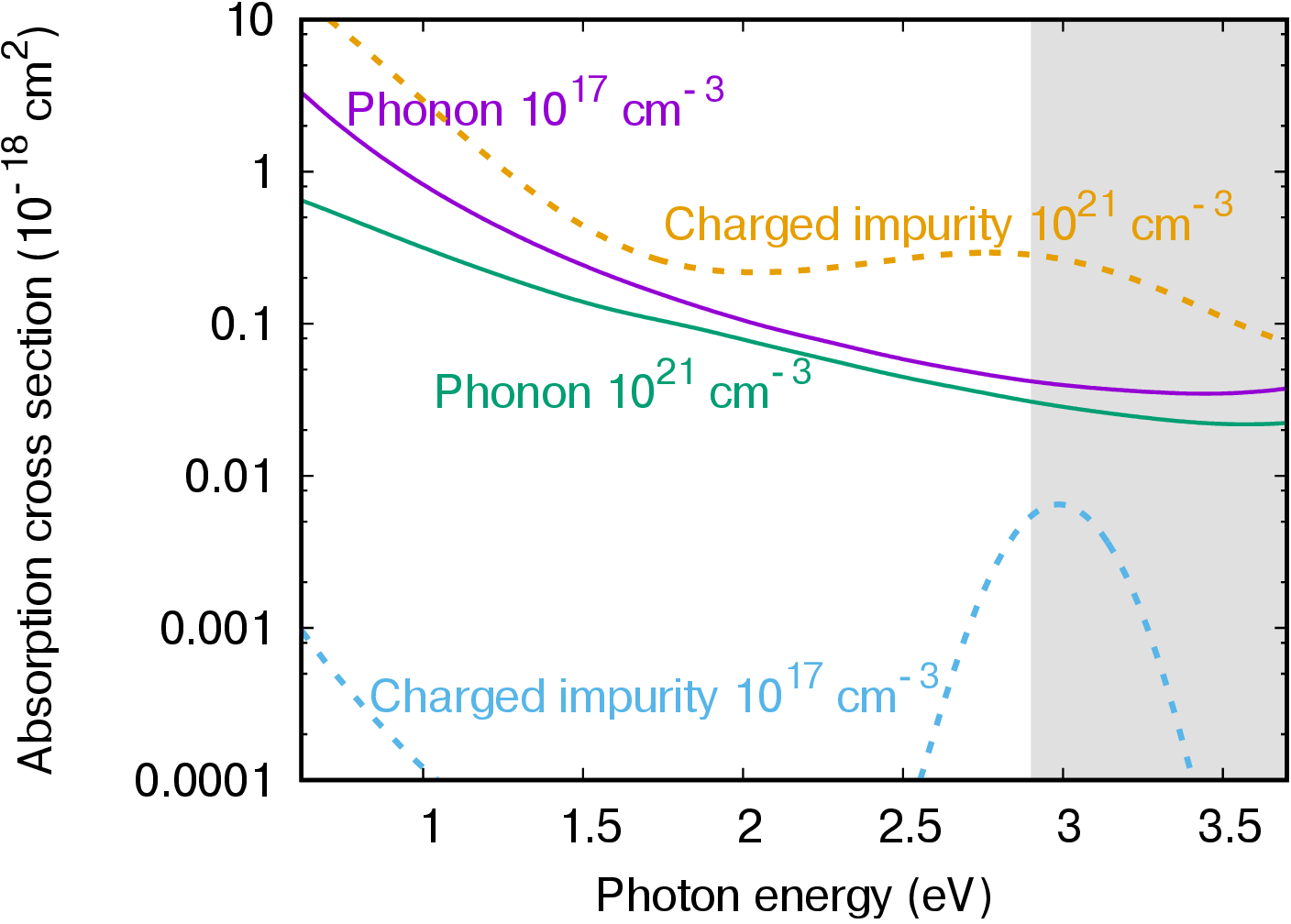}
	\caption{(Color online) Phonon- (solid lines) and charged-impurity-assisted (dashed lines) absorption cross section for In$_2$O$_3$ as a function of photon energy, for two representative concentrations of free carriers. Screening is included as described in the text. }
	\label{fig:comparison}
\end{figure}

At high carrier concentrations, screening of the electron-phonon interaction can be significant.  This effect is included by multiplying the electron-phonon matrix elements  by $\frac{q^2}{q^2+q^2_{\text{scr}}}$.
$q_{\text{scr}}$ is a screening length, which is obtained from either the Debye model (non-degenerate case) or the Thomas-Fermi model (degenerate case).
Results are shown in Fig.~\ref{fig:comparison} for two values of the electron concentration.
The curve labeled ``Phonon 10$^{17}$ cm$^{-3}$'' in Fig.~\ref{fig:comparison} is practically equivalent to the unscreened result (shown previously in Fig.~\ref{fig:indabs}), since for these low carrier concentrations the effect of screening is negligible.
At higher free-carrier concentrations, screening becomes important, reducing the phonon-assisted absorption cross section. The screening effect is larger for smaller photon energies.

We now turn to a discussion of free-carrier absorption assisted by a different mechanism, namely charged-impurity scattering.  This process is unavoidable, because the high conductivity of a TCO requires doping, which leads to the presence of ionized defects or impurities.
In In$_2$O$_3$, the free carriers are typically provided by Sn doping, leading to the presence of charged Sn$^+$ centers.
We describe the resulting indirect absorption process by scattering off a screened Coulomb potential, where the matrix elements are given by
\begin{equation}
g^{\text{impurity}}_{i,j}=\left< i,\bm{k}\middle|\frac{4\pi e^2 Z}{\epsilon_0 (q^2+q^2_{\text{scr}})}  \middle| j,\bm{k}+\bm{q} \right>.
\end{equation}
We assume here that all Sn dopants are ionized and that these are the only charged impurities present; the concentration of ions is then equal to the concentration of free carriers.
This assumption neglects compensation and thus provides a lower limit for the total concentration of charged centers.

As seen in Fig.~\ref{fig:comparison}, at 10$^{17}$ cm$^{-3}$, charged-impurity-assisted absorption is negligible compared to the phonon-assisted process, but
at 10$^{21}$ cm$^{-3}$, charged-impurity scattering dominates. For photon energies up to 2 eV the absorption decreases with increasing photon energies, similar to the phonon-assisted process. Above 2 eV, the impurity-assisted absorption increases, with a peak around 3 eV.
As mentioned before, for energies around 2 eV above the conduction-band minimum, the first conduction band becomes degenerate with the second conduction band (Fig.~\ref{fig:bandplot}). The corresponding matrix elements are larger for the charged-impurity process compared to the phonon process, leading to an increase in absorption. In addition, around 3 eV, many more final states become available (see, e.g., the band crossings at the H point in Fig.~\ref{fig:bandplot}). Note that in the band structure obtained using hybrid functionals (see Fig. S1)  the higher conduction bands are located at slightly higher energies compared to our LDA calculations; this will move the aforementioned peaks to higher energies.
The crossover point, where the impurity-assisted process is similar in magnitude to the phonon-assisted process, occurs at an impurity concentration around 3$\times$10$^{20}$ cm$^{-3}$; at higher doping concentrations impurity-assisted process becomes the dominant absorption process.

In conclusion, we have reported a detailed analysis of the limitations on transparency due to indirect free-carrier absorption in In$_2$O$_3$.
Using first-principles techniques, we find that the phonon-assisted process is dominated by {\it emission} of phonons, and increases with decreasing photon energies. This increase can be described by a third power dependence on the wavelength. For long-wavelength photons, a Fr\"ohlich model, only taking the highest LO phonon mode into account, provides a reasonable description of the indirect absorption up to about 2.5 eV.
This model can also explain why phonon-assisted indirect absorption in In$_2$O$_3$ is about 50\% smaller than this in SnO$_2$.
Charged-impurity-assisted absorption becomes the dominant process for concentrations above 3$\times$10$^{20}$ cm$^{-3}$.

\section*{Supplementary Material}
See supplementary material for a comparison of the band structure calculated with LDA and with a hybrid functional, a comparison of the calculated Raman-active phonon frequencies with previous calculations and experiments, and a comparison of the phonon-assisted indirect absorption at 0 K and 300 K.

\section*{Acknowledgments}
This work was supported by the GAME MURI of the Air Force Office of Scientific Research (FA9550-18-1-0479).
Computational resources where provided by the Extreme Science and Engineering Discovery Environment (XSEDE), which is supported by National Science Foundation grant number ACI-1548562.

%
\end{document}


\title{Supplementary Material: Limitations of In$_2$O$_3$ as a transparent conducting oxide}
\author{H. Peelaers}
  \email{peelaers@ku.edu}
  \affiliation{Materials Department, University of California, Santa Barbara, California 93106-5050, USA}
  \affiliation{Department of Physics and Astronomy, University of Kansas, Lawrence, Kansas 66045, USA}
\author{E. Kioupakis}
\affiliation{Department of Materials Science and Engineering, University of Michigan, Ann Arbor, Michigan 48109, USA}
\author{C. G. Van de Walle}
\affiliation{Materials Department, University of California, Santa Barbara, California 93106-5050, USA}

\maketitle
\renewcommand{\thetable}{S\arabic{table}}
\renewcommand{\thefigure}{S\arabic{figure}}

\section{Comparison of band structures calculated with LDA and with a hybrid functional}

For the comparison of the band structure we used fully relaxed calculations using the HSE06~\cite{Heyd2003,Heyd2006} hybrid functional. To be able to directly compare the LDA band structure and the HSE06 band structure, we set the valence-band maxima to zero, and used a scissor shift so that the conduction-band minima occur at 2.9 eV. Fig. S1 shows this comparison, where solid purple lines are the LDA results, and dashed black lines the HSE06 results. The curvature of the lowest conduction band is slightly larger for LDA compared to HSE06. To quantify this, we calculated the effective mass, obtaining 0.16 m$_e$ for LDA and 0.21 m$_e$ for HSE.  The experimental value is 0.18  $\pm$ 0.02 m$_e$~\cite{Feneberg2016}, showing that the LDA value agrees with experiment to within the experimental error bar. The higher conduction bands are located at slightly higher energies for the HSE06 band structure compared to the LDA band structure.
\begin{figure}[bh!]
\includegraphics[width=0.8\columnwidth]{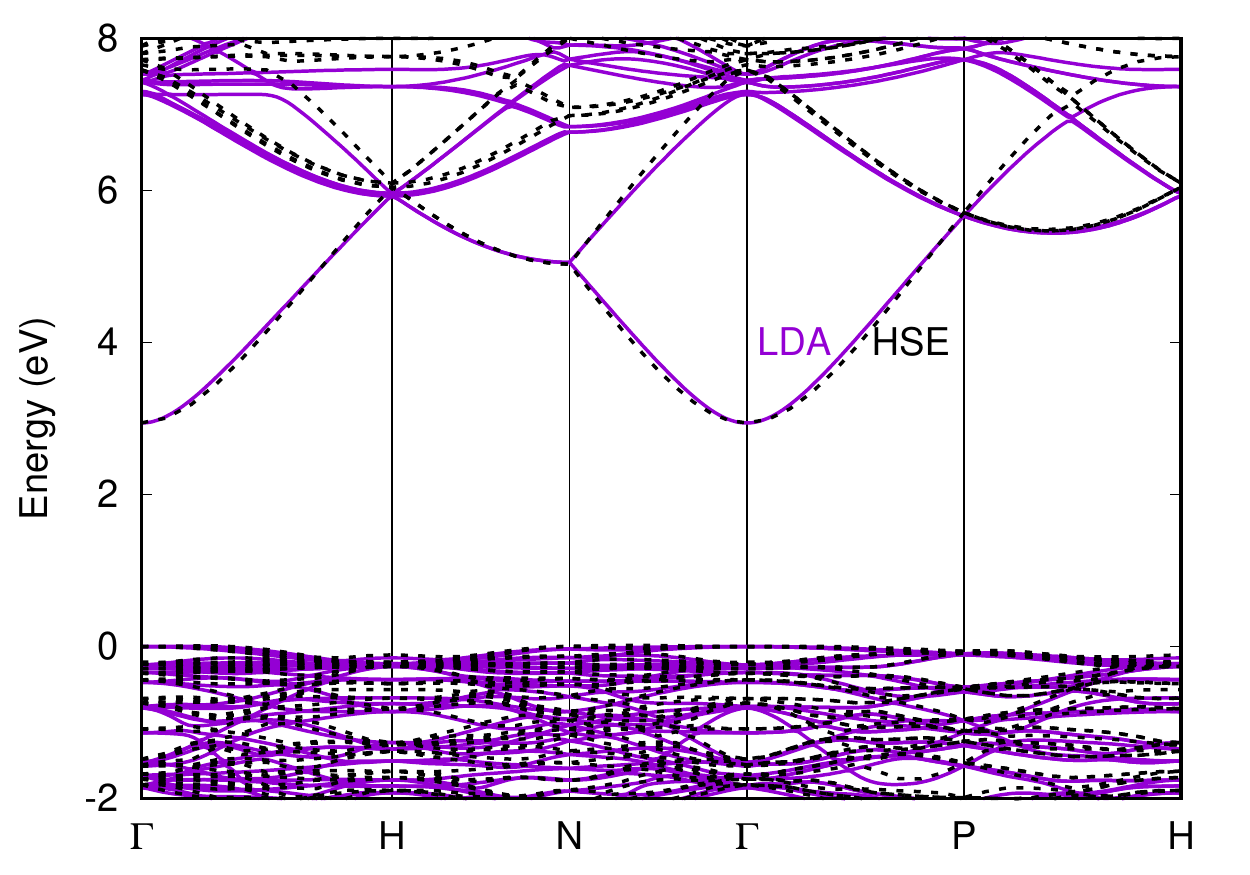}
\caption{Calculated band structure of In$_2$O$_3$ using LDA (solid purple lines) and HSE06 (dashed black lines). The valence-band maxima of both band structures are set to 0, and the conduction-band minima to 2.9 eV.}
\end{figure}
\newpage

\section{Comparison of phonon frequencies}
\begin{table*}[h]
\centering
\caption{Calculated phonon fequencies (in cm$^{-1}$) for Raman-active phonon modes in In$_2$O$_3$ and comparison with previous calculations~\cite{Garcia-Domene2012} and experiments~\cite{Garcia-Domene2012,Kranert2014,White1972,Korotcenkov2005,MateiGhimbeu2008,Zhang2007,Berengue2010,Rojas-Lopez2000}.}
\begin{ruledtabular}
  \begin{tabular}{ccccccccccc}
Mode & This work & Ref.~\onlinecite{Garcia-Domene2012} & Ref.~\onlinecite{Garcia-Domene2012} & Ref.~\onlinecite{Kranert2014} & Ref.~\onlinecite{White1972} & Ref.~\onlinecite{Korotcenkov2005} & Ref.~\onlinecite{MateiGhimbeu2008} & Ref.~\onlinecite{Zhang2007} & Ref.~\onlinecite{Berengue2010} & Ref.~\onlinecite{Rojas-Lopez2000} \\
Symmetry &  & calculated &  &  &  & &  & & &  \\
\hline
  T$_g$                                                      & 100       & 106                           & 108                    &             &           &                 &             &           & 109          &                 \\
  T$_g$                                                      & 111       & 114                           & 118                    & 112         &           &                 &             &           &              &                 \\
  A$_g$                                                      & 130       & 128                           & 131                    & 135         &           &                 & 135         & 131       & 135          & 131             \\
  T$_g$                                                      & 146       & 148                           & 152                    &             &           &                 &             &           &              &                 \\
  E$_g$                                                      & 165       & 165                           & 169                    & 172         &           &                 &             &           &              &                 \\
  T$_g$                                                      & 205       & 199                           & 205                    & 215         &           &                 &             &           &              &                 \\
  T$_g$                                                      & 207       & 204                           & 211                    & 219         &           &                 &             &           &              &                 \\
  T$_g$                                                      & 312       & 302                           & 306                    & 307         & 308       & 306             & 307         & 306       & 307          &                 \\
  A$_g$                                                      & 318       & 302                           & 306                    & 310         &           &                 &             &           &              & 303             \\
  E$_g$                                                      & 320       & 308                           &                        & 317         &           &                 &             &           &              &                 \\
  T$_g$                                                      & 324       & 312                           &                        &             &           &                 &             &           &              &                 \\
  T$_g$                                                      & 373       & 356                           & 365                    & 369         & 365       & 366             & 365         & 365       & 366          &                 \\
  T$_g$                                                      & 396       & 379                           &                        & 392         &           &                 &             &           &              &                 \\
  E$_g$                                                      & 401       & 385                           & 396                    & 400         &           &                 &             &           &              &                 \\
  T$_g$                                                      & 448       & 438                           &                        & 454         &           &                 &             &           &              &                 \\
  T$_g$                                                      & 465       & 447                           & 467                    & 469         & 471       &                 &             &           &              &                 \\
  A$_g$                                                      & 487       & 476                           & 495                    & 499         & 504       & 495             & 496         & 495       & 495          & 496             \\
  T$_g$                                                      & 516       & 499                           &                        & 520         &           &                 &             &           & 517          &                 \\
  T$_g$                                                      & 538       & 520                           &                        & 547         &           &                 &             &           &              &                 \\
  E$_g$                                                      & 584       & 565                           & 590                    & 589         &           &                 &             &           &              &                 \\
  A$_g$                                                      & 595       & 576                           &                        & 629         &           &                 &             & 602       &              & 602             \\
  T$_g$                                                      & 618       & 600                           & 628                    & 633         & 637       & 630             & 627         & 628       & 630          & \\
\end{tabular}
\end{ruledtabular}
\end{table*}
\newpage

\section{Temperature dependence of the indirect phonon-assisted absorption}
\begin{figure}[tbh!]
\includegraphics[width=0.8\columnwidth]{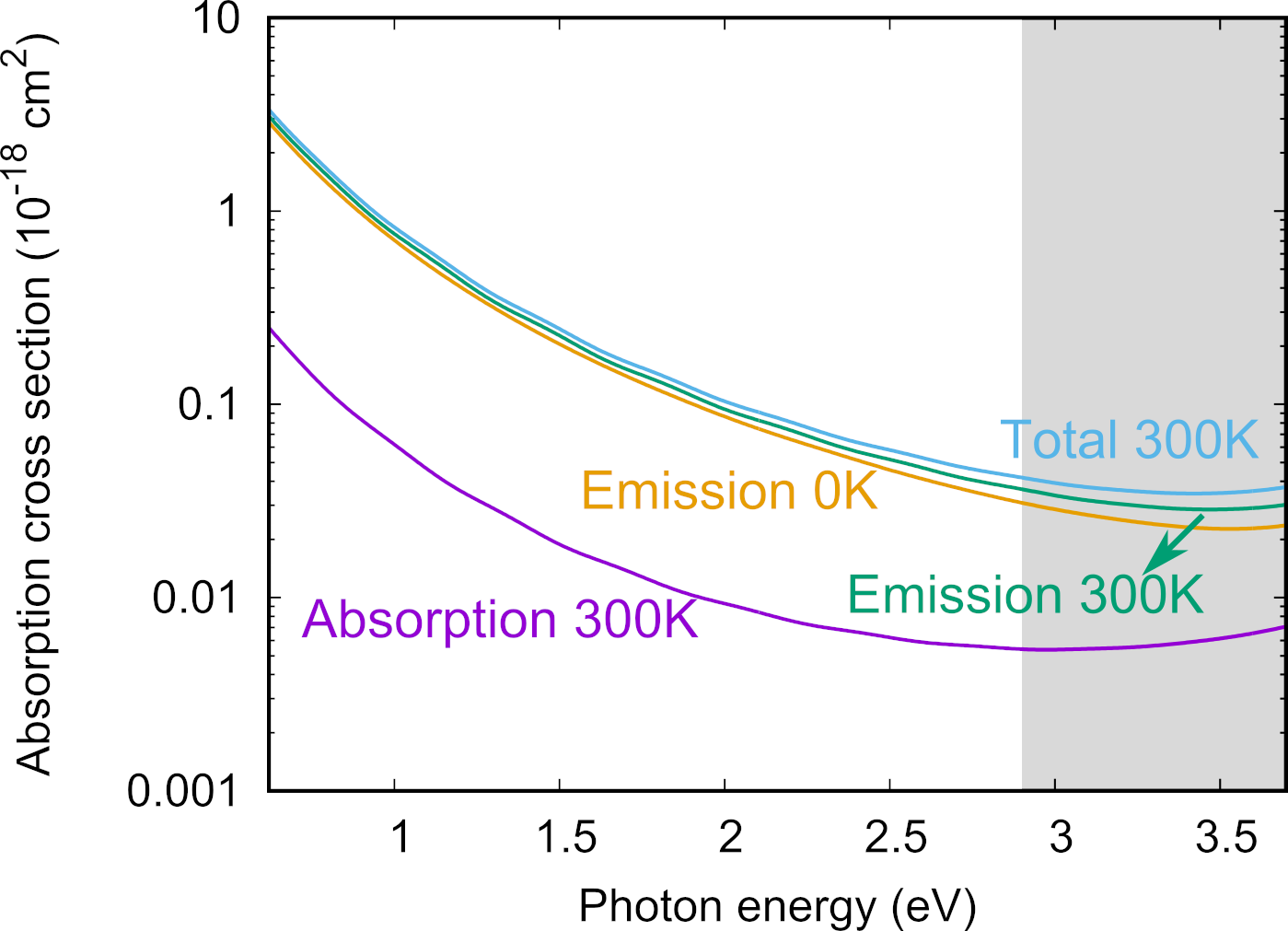}
\caption{Calculated phonon-assisted absorption cross section for In$_2$O$_3$ as a function of photon energy for $T$=300 K and $T$=0 K. The grey area indicates energies for which weak across-the-gap transitions can take place. Contributions from phonon absorption {\it versus} phonon emission are identified. At 0 K, there is no contribution from phonon absorption, so that the total cross section is equal to the contribution from phonon emission.}
\end{figure}

%